\documentstyle[12pt]{article}
\input{psfig}
\def\Journal#1#2#3#4{{#1} {\bf #2} (#4) #3}

\def\NPA{{\rm Nucl. Phys.} A}
\def\NPB{{\rm Nucl. Phys.} B}
\def\PLB{{\rm Phys. Lett.}  B}
\def\PRL{\rm Phys. Rev. Lett.}
\def\PRD{{\rm Phys. Rev.} D}

\def\ZPC{{\rm Z. Phys.} C}
\def\PAN{\rm Phys. At. Nuclei }
\def\la{\langle}
\def\ra{\rangle}
\begin{document}
\title{ Light-Front Quark Model Analysis of Exclusive 
$0^{-}$$\to$$0^{-}$ Semileptonic Heavy Meson Decays}
\author{ Ho-Meoyng Choi$^{(a)}$\thanks{hmchoi@unity.ncsu.edu}
and Chueng-Ryong Ji$^{(a,b)}$\thanks{ji@ncsu.edu} \\  
$^{(a)}$Department of Physics, North Carolina State University \\
Raleigh, N.C. 27695-8202, USA \\
$^{(b)}$Asia Pacific Center for Theoretical Physics, Seoul 130-01, Korea}
\date{\empty}
\maketitle
\begin{abstract}
We present the analysis of exclusive $0^{-}$$\to$$0^{-}$ semileptonic
heavy meson decays using the constituent quark model based
on the light-front quantization.
Our model is constrained by the variational principle for the well-known
linear plus Coulomb interaction motivated by QCD.
Our method of analytic continuation to obtain the weak form factors avoids
the difficulty associated with the contribution from the nonvalence
quark-antiquark pair creation. Our numerical results for the heavy-to-heavy
and heavy-to-light meson decays are in a good agreement with the
available experimental data and the lattice QCD results.
In addition, our model predicts the two unmeasured mass spectra of
$^{1}S_{0}(b\bar{b})$ and $^{3}S_{1}(b\bar{s})$ systems as
$M_{b\bar{b}}$=9657 MeV and $M_{b\bar{s}}$= 5424 MeV.
\end{abstract}
\newpage
\baselineskip=20pt
In recent years, the exclusive semileptonic decay processes generated a 
great excitement not only in measuring the most accurate values of the 
Cabbibo-Kobayashi-Maskawa (CKM) matrix elements but also in testing diverse
theoretical approaches to describe the internal structure of hadrons.
Especially, due to the anticipated abundance of accurate experimental data
from the $B$-factories (e.g. HERA-B at HERA, BaBar at SLAC and Belle at KEK), 
the heavy-to-heavy and heavy-to-light meson
decays such as $B$$\to$$D$, $B$$\to$$\pi$, $D$$\to$$\pi(K)$ etc. 
become invaluable processes deserving thorough analysis.
While the available experimental data of heavy meson branching ratios 
have still rather large uncertainties\cite{data}, various theoretical methods 
have been applied to calculate the weak decay processes, e.g., lattice 
QCD\cite{Flynn,UKQCD,Bernard}, QCD sum rules\cite{Ball}, Heavy quark
effective theory\cite{IW}, 
and quark models\cite{CJ1,CJ2,Isgur,Wirbel,Dem,Mel,Jaus}.
In particular, the weak transition form factors determined by the lattice 
QCD\cite{Bernard} provided a useful guidance for the model building of hadrons, 
making definitive tests on existing models, even though the current error bars 
in the lattice data are yet too large to pin down the best phenomenological 
model of hadrons. These weak form factors, however, are the essential 
informations of the strongly interacting quark/gluon structure inside hadrons
and thus it is very important to analyze these processes with the viable
model that has been very successful in analyzing other processes.

In this letter, we report the analysis of exclusive semileptonic decays 
of a heavy 
pseudoscalar meson into another heavy or light pseudoscalar meson using 
the light-front quark model (LFQM) which has been quite successful in the 
analysis of electromagnetic form factors, radiative decays and 
$K$$\to$$\pi$ transition form factors\cite{CJ1,CJ2}.  
The LFQM takes advantage of the equal LF time ($\tau$=$t+z/c$) 
quantization\cite{Brod1} and includes important relativistic effects 
in the hadronic wave functions. The distinguished feature of the LF 
equal-$\tau$ quantization compared to the ordinary equal-$t$ quantization is
the rational energy-momentum dispersion relation\cite{CJ3} which leads to the 
suppression of vacuum fluctuations with the decoupling of complicated 
zero modes\cite{CJ4,Ji} and the conversion of the dynamical problem from boost 
to rotation\cite{Ji2}. The recent lattice QCD results\cite{Kura}
indicated that the mass difference
between $\eta'$ and pseudoscalar octet mesons due to the complicated nontrivial
vacuum effect increases (or decreases) as the quark mass $m_{q}$ decreases 
(or increases), i.e., the effect of the topological charge contribution should
be small as $m_{q}$ increases. This supported in building the constituent quark
model\cite{CJ1} in the LF quantization approach because the complicated 
nontrivial vacuum effect in QCD can be traded off by the rather large 
contstituent quark masses. We have also circumvented the problem of assigning
the dynamical quantum numbers $J^{PC}$ to hadrons by using the Melosh 
transformation of each constituent from equal $t$ to equal $\tau$\cite{CJ1}. 

Moreover, one of the most distinctive advantages in the  
LFQM has been the utility of the well-established Drell-Yan-West 
($q^{+}$=$q^{0}+q^{3}$=0) frame for the calculation of various form 
factors\cite{Brod2}. 
By taking the ``good" components of the current ($j^{+}$ and ${j}_{\perp}$), 
one can get rid of the zero mode\cite{CJ4} problem and compute the full 
theoretical prediction for
the spacelike form factors in $q^{+}$=0 frame. The weak
transition form factors that we are considering, however, are the 
timelike $q^{2}>0$ observables. 
Our method is to rely on the analytic continuation
from the spacelike region to the timelike region calculating the ``good"
components of the current in the $q^{+}$=0 frame\cite{CJ2}.
If we were to take the $q^{+}$$\neq$0 frame, then we must take 
into account the higher Fock-state (nonvalence) contributions 
arising from quark-antiquark pair creation (so called ``Z-graph") as well as 
the valence configuations. In fact, we notice that a few previous 
analyses\cite{Dem} were performed in the $q^{+}$$\neq$0 frame without taking 
into account the nonvalence contributions. 
We find that such omission leads to a large deviation from the full results 
\cite{CJ2}. Our method is to rely on the analytic continuation
from the spacelike region to the timelike region calculating the ``good"
components of the current in the $q^{+}$=0 frame. 

The key idea in our LFQM\cite{CJ1} for mesons is to treat
the radial wave function as a trial function for the variational
principle to the QCD-motivated Hamiltonian saturating 
the Fock state expansion by the \underline {constituent} quark and antiquark. 
The spin-orbit wave function is uniquely determined by the Melosh 
transformation. 
We take the QCD-motivated effective Hamiltonian as the well-known linear plus
Coulomb interaction given by
\begin{eqnarray}
H_{q\bar{q}}= H_{0} + V_{q\bar{q}} 
=\sqrt{m_{q}^{2}+k^{2}} + \sqrt{m_{\bar{q}}^{2}+k^{2}}+ V_{q\bar{q}},
\end{eqnarray}
where 
\begin{eqnarray}
V_{q\bar{q}}= V_{0} + V_{\rm hyp}
= a + br - \frac{4\kappa}{3r}
+ \frac{2\vec{S}_{q}\cdot\vec{S}_{\bar{q}}}
{3m_{q}m_{\bar{q}}}\nabla^{2}V_{\rm Coul}.
\end{eqnarray}
We take the Gaussian radial wave function 
$\phi(k^{2})=N\exp(-k^{2}/2\beta^{2})$ 
as our trial wave function\footnote{ Even though one
can in principle expand the radial function $\phi_{n,l=0}(k^{2})$ with
a truncated set of HO basis states\cite{Isgur}, our choice of radial
wave function turns out to be sufficient for the analysis of
the ground state $0^{-+}$ and $1^{--}$ ground state meson 
properties\cite{CJ1}.}
to minimize the central Hamiltonian\cite{CJ1}.
Among the light-quark mass and the potential parameters
$\{m_{u}(=m_{d}), \beta_{u\bar{d}}(=\beta_{u\bar{u}}), a, b, \kappa\}$,  
only 4 parameters 
are independent 
because of the constraint from the variational principle.
Furthermore, the string tension 
$b$=0.18 GeV$^{2}$ and the constituent $u$ and $d$ quark masses 
$m_{u}$=$m_{d}$=0.22 GeV are rather well known from other quark model 
analyses commensurate with Regge phenomenology\cite{Isgur}.  
Thus, using the experimental values of $\rho$ and $\pi$ masses
and the variational constraint, we can fix the remaining
parameters $a,\kappa$, 
and $\beta_{u\bar{d}}$ as $a=-0.724$ GeV, 
$\beta_{u\bar{d}}=0.3695$ GeV, and $\kappa=0.313$, respectively\cite{CJ1}.  
More detailed procedure of determining the model parameters of light-quark
sector ($u$ and $s$) can be found in \cite{CJ1}.
It is very important to note that 
all other model parameters such as $m_{c}$, $m_{b}$, 
$\beta_{uc}$, $\beta_{ub}$, etc. are then uniquely determined by the same 
procedure as the light-quark analysis \cite{CJ1}. 
The procedure of determining model parameters 
constrained by the variational principle \cite{CJ1} is shown
in Fig. 1, where the lines of $qq$ and $qc$ ($q$=$u$ and $d$) etc. 
represent the sets of $\{m_{q},m_{q},\beta_{qq}\}$ and 
$\{m_{q},m_{c},\beta_{qc}\}$, respectively, etc.  
Because all the lines in Fig. 1 should go through the same point of 
($b$=0.18 GeV$^{2}$,$\kappa=0.313$), the parameters of $m_{c}$, $m_{b}$,
$\beta_{uc}$, $\beta_{ud}$, etc. are all automatically determined  without 
any adjustment. Our model parameters obtained by the variational principle 
are summarized in Table 1.

Our predictions of the ground state meson mass spectra and the decay constants 
of various heavy pseudoscalar mesons are summarized in Tables 2 and 3, 
respectively, and compared with the available experimental data\cite{data} and  
the lattice QCD results\cite{Flynn}. Our predictions of ground state meson 
mass spectra agree with the experimental data\cite{data} 
within 6$\%$ error. Furthermore, our model predicts the two unmeasured mass 
spectra of $^{1}S_{0}(b\bar{b})$ and $^{3}S_{1}(b\bar{s})$ systems as 
$M_{b\bar{b}}$=9657 MeV and $M_{b\bar{s}}$=5424 MeV, respectively.
Our values of the decay constants are also in a good 
agreement with the results of lattice QCD\cite{Flynn} anticipating 
future accurate experimental data. 

The matrix element of the current $j^{\mu}=\bar{q}_{2}\gamma^{\mu}Q_{1}$ for  
$0^{-}(Q_{1}\bar{q})$$\to$$0^{-}(q_{2}\bar{q})$ decay is given by  
two weak form factors $f_{+}$ and $f_{-}$, viz.,  
\begin{eqnarray}
\la P_{2}|\bar{q_{2}}\gamma^{\mu}Q_{1}|P_{1}\ra=
f_{+}(q^{2})(P_{1}+P_{2})^{\mu} + f_{-}(q^{2})q^{\mu},  
\end{eqnarray}
where $q^{\mu}=(P_{1}-P_{2})^{\mu}$ is the four-momentum transfer to
the lepton and $m^{2}_{\ell}\leq q^{2}\leq (M_{1}-M_{2})^{2}$. 
In the heavy quark limit $M_{1(2)}$$\to$$\infty$, the form factor 
$f_{+}(q^{2})$ is reduced to the universal Isgur-Wise (IW) 
function, $\xi(v_{1}\cdot v_{2})= [2\sqrt{M_{1}M_{2}}/ (M_{1}+M_{2})]
f_{+}(q^{2})$, where $v_{1(2)}=P_{1(2)}/M_{1(2)}$.   
In LFQM, the matrix element of the weak vector 
current  can be obtained by the convolution of initial and final 
LF meson wave functions in $q^{+}$=0 frame: 
\begin{eqnarray}
& &\la P_{2}|\bar{q}_{2}\gamma^{\mu}Q_{1}|P_{1}\ra\nonumber\\ 
& &= -\int^{1}_{0}dx\int d^{2}{\bf k}_{\perp}
\frac{\phi^{\dagger}_{2}(x,{\bf k'}_{\perp})\phi_{1}(x,{\bf k}_{\perp})}
{2(1-x)\prod^{2}_{i=1}\sqrt{ M^{2}_{i0} -(m_{i}-m_{\bar{q}})^{2} }}
\nonumber\\
& &\;\;\;\;\times{\rm Tr} \biggl[\gamma_{5}({\not\! p}_{2}+m_{2})
\gamma^{\mu}({\not\!p}_{1}+m_{1})\gamma_{5}
({\not\! p}_{\bar{q}}-m_{\bar{q}})\biggr],
\end{eqnarray}
where $M^{2}_{i0} = (k_{\perp}^{2}+m^{2}_{i})/(1-x)
+ (k_{\perp}^{2}+m^{2}_{\bar{q}})/x$. 
In Eq. (4), the form factors $f_{+}(q^{2})$ and $f_{-}(q^{2})$ 
are obtained by taking the ``good" components of the current  
($j^{+}$ and $j_{\perp}$). The detailed
derivation of $f_{\pm}(q^{2})$ can be found in \cite{CJ2}.
Our analytic continuation to the timelike region has verified\cite{CJ2} 
the equivalence to the dispersion method\cite{Mel}. 

Our numerical results of the decay rates for
$D$$\to$$\pi(K)$, $D_{s}$$\to$$\eta(\eta')$, and $B\to\pi(D)$ processes
are consistent with the experimental data as summarized in Table 4.
It is interesting to note that our value of $\eta$-$\eta'$ mixing angle,
$\theta_{SU(3)}$=$-19^{\circ}$ presented in \cite{CJ1}, are also in 
agreement with the data for $D_{s}$$\to$$\eta(\eta')$ decays.  
One should note that the number of events for the $D\to\pi$ data is 
currently very small compared to other processes\cite{data}.

In Figs. 2(a) and 2(b), we present the
form factors $f^{DK}_{+}(q^{2})$ and $f^{DK}_{0}(q^{2})$, respectively,
with the definition of $f_{0}(q^{2})$ as 
$f_{0}(q^{2})=f_{+}(q^{2})+q^{2}f_{-}(q^{2})/(M^{2}_{1}-M^{2}_{2})$, 
and compare with the experimental data as well as the  
lattice QCD results\cite{Bernard}.
Note that our value of $f^{DK}_{+}(0)$=0.736
is within the error bar of the measured value\cite{data},
$f^{\rm Expt.}_{+}(0)=0.7\pm0.1$.
In Fig. 3, we present the form factor $f^{B\pi}_{+}(q^{2})$ and
compare with the results from lattice QCD\cite{UKQCD}. 
Our result is very close to the UKQCD\cite{UKQCD}
results for a wide range of momentum transfer.
In Fig. 4, our prediction of the IW function for $B$$\to$$D$ transition
are compared with the experimental data\cite{Argus,CLEO}.
Our prediction of the slope $\rho^{2}$=0.8 of the IW function at the
zero-recoil point defined as $\xi(v_{1}\cdot v_{2})= 1-\rho^{2}(v_{1}\cdot
v_{2}-1)$ is quite comparable with the current world average 
$\rho_{\rm avg.}$=0.66$\pm$0.19\cite{data} extracted from exclusive 
semileptonic $\bar{B}$$\to$$D\ell\bar{\nu}$ decay.  

In conclusion, in this paper, we analyzed the exclusive 
$0^{-}$$\to$$0^{-}$ 
semileptonic heavy meson decays using the LFQM constrained by the variational
principle for the QCD-motivated effective Hamiltonian with the well-known
linear plus Coulomb interaction. 
Our model not only provided overall a good agreement with the available
experimental data and the lattice QCD results for the weak transition 
form factors and branching ratios of the heavy-to-heavy and heavy-to-light
meson decays but also rendered a large number of predictions to the heavy
meson mass spectra and decay constants. 
Our model in fact predicted the masses of 
heavy mesons, i.e., $M_{b\bar{b}}(^{1}S_{0})$=9657 MeV and 
$M_{b\bar{s}}(^{3}S_{1})$= 5424 MeV. 
We have overcome the difficulty associated with the nonvalence Z-graph 
contribution in timelike region by the analytic continuation of weak form
factors from the spacelike region. 
Our numerical computation confirmed the equivalence of our analytic 
continuation method and the dispersion relation 
method\cite{Mel}\footnote{If we were to use the model parameters given in 
\cite{Mel}, we obtain the values of $f_{+}(0)-f_{+}(q^{2}_{\rm max})$ as 
follows: 0.783 (0.781) - 1.2 (1.2) for $D\to K$, 0.682 (0.681) - 1.61 (1.63)
for $D\to\pi$, 0.682 (0.684) - 1.12 (1.12) for $B\to D$, and 
0.293 (0.293) - 2.7 (2.3) for $B\to\pi$, where the values in parentheses are the
results obtained by the auther in \cite{Mel}.}. 
We think that the success of our model hinges on the advantage of light-front
quantization realized by the rational energy-momentum dispersion relation.  It is 
crucial to calculate the ``good" components of the current in the reference frame 
which deletes the complication from the nonvalence Z-graph contribution. 
We anticipate further stringent tests of our model with more accurate data 
from future experiments and lattice QCD calculations.

This work was supported by the Department of Energy under Grant No. 
DE-FG02-96ER40947. The North Carolina Supercomputing Center and the National
Energy Research Scientific Computer Center are also acknowledged for the 
grant of supercomputer time.
 

\clearpage 
\begin{table}
\caption{The constituent quark masses $m$[GeV]
and the Gaussian parameters $\beta$[GeV] for the  
the linear potential obtained by the variational principle. 
$q$=$u$ and $d$.}
\begin{center}
{\footnotesize 
\begin{tabular}{|c|c|c|c|c|c|c|c|c|c|c|c|c|}
\hline
$m_{q}$ & $m_{s}$ & $m_{c}$ & $m_{b}$ & $\beta_{q\bar{q}}$ &
$\beta_{s\bar{s}}$& $\beta_{q\bar{s}}$ & $\beta_{q\bar{c}}$ &
$\beta_{s\bar{c}}$ & $\beta_{c\bar{c}}$& $\beta_{q\bar{b}}$&
$\beta_{s\bar{b}}$ & $\beta_{b\bar{b}}$ \\
\hline 
0.22 & 0.45 & 1.8 & 5.2 &0.3659 & 0.4128 & 0.3886
& 0.4679 & 0.5016 & 0.6509 & 0.5266 & 0.5712 & 1.1452 \\
\hline
\end{tabular}
}
\end{center}
\end{table}

\begin{table}
\caption{Fit of the ground state meson masses [MeV]  with
the parameters given in Table I. Underline masses are input data.
The masses of ($\omega-\phi$) and ($\eta-\eta'$) were used to
determine the mixing angles of $\omega-\phi$ and $\eta-\eta'$ [7],
respectively.}
\begin{center}
\begin{tabular}{|c|c|c|c|c|c|} \hline 
$^{1}S_{0}$ & Expt.\cite{data}  & Prediction  &
$^{3}S_{1}$ & Expt.\cite{data} & Prediction  \\
\hline
$\pi$       & 135 $\pm$0.0006 & \underline{135} &
$\rho$      & 770 $\pm$ 0.8   & \underline{770}\\
$K$         & 498 $\pm$ 0.016 & 478 &
$K^{*}$     & 892 $\pm$ 0.26  & 850             \\
$\eta$      & 547 $\pm$ 0.12  & \underline{547} &
$\omega$    & 782 $\pm$ 0.12  & \underline{782}\\
$\eta'$     & 958 $\pm$0.14   & \underline{958} &
$\phi$      & 1020$\pm$0.008  & \underline{1020}\\
$D$         & 1865$\pm$0.5    & 1836  &
$D^{*}$     & 2007$\pm$ 0.5   & 1998 \\
$D_{s}$     & 1969$\pm$0.6    & 2011  &
$D^{*}_{s}$ & 2112$\pm$0.7    & 2109  \\
$\eta_{c}$  & 2980$\pm$2.1    & 3171 &
$J/\psi$    & 3097$\pm$0.04   & 3225 \\
$B$         & 5279$\pm$ 1.8   & 5235 &
$B^{*}$     & 5325$\pm$ 1.8   & 5315 \\
$B_{s}$     & 5369$\pm$2.0    & 5375 &
$(b\bar{s})$& --              & 5424 \\
$(b\bar{b})$& --              & 9657  &
$\Upsilon$  & 9460$\pm$ 0.21  & 9691 \\
\hline
\end{tabular}
\end{center}
\end{table}

\clearpage 
\begin{table}
\caption{Decay constants[MeV] for various heavy pseudoscalar mesons.}
\begin{center}
\begin{tabular}{|c|c|c|c|c|} \hline
References& $f_{D}$ & $f_{D_{s}}$& $f_{B}$& $f_{B_{s}}$\\
\hline
Ours  & 139.2 & 164.8 &121.2  & 144.2 \\
\hline
Lattice\cite{Flynn}& 141.4$\pm$21.2 & 155.6$\pm$21.2 &
120.2$\pm$24.7 & 137.9$\pm$24.7\\
\hline
Expt.\cite{data} & $<219$ & $137-304$ & -- & --\\
\hline
\end{tabular}
\end{center}
\end{table}
\begin{table}
\begin{center}
\caption{ Form factors $f_{+}(0)$ and branching ratios (Br.) for various 
heavy meson semileptonic decays for $0^{-}$$\to$$0^{-}$. 
We use $\theta^{\eta-\eta'}_{SU(3)}$=$-19^{\circ}$ for 
$D_{s}\to\eta(\eta')$ decays and the following CKM matrix element:
$|V_{cs}|$=1.04$\pm$0.16, 
$|V_{cd}|$=0.224$\pm$0.016, $|V_{ub}|$=(3.3$\pm$0.4$\pm$0.7)$\times$10$^{-3}$, 
and $|V_{bc}|$=0.0395$\pm$0.003 [1].}
\begin{tabular}{|c|c|c|c|}\hline
Processes & $f_{+}(0)$ & Br. & Expt.\cite{data} \\ \hline
$D\to K$  & 0.736 & $(3.75\pm 1.16)\%$ & $(3.66\pm0.18)\%$ \\ \hline 
$D\to\pi$ & 0.618 & $(2.36\pm 0.34)\times 10^{-3}$
          & $(3.9^{+2.3}_{-1.1}\pm 0.4)\times 10^{-3}$ \\ \hline
$D_{s}\to\eta$ & 0.421 & $(1.8\pm 0.6)\%$ & $(2.5\pm 0.7)\%$ \\ \hline
$D_{s}\to\eta'$ & 0.585 & $(9.3\pm 2.9)\times 10^{-3}$ 
          & $(8.8\pm 3.4)\times 10^{-3}$ \\ \hline 
$B\to\pi$ & 0.273 & $(1.40\pm 0.34)\times 10^{-4}$ & 
$(1.8\pm0.6)\times 10^{-4}$ \\ \hline 
$B\to D$  & 0.709 & $(2.28\pm 0.20)\%$ & $(2.00\pm0.25)\% $\\ 
\hline
\end{tabular}
\end{center}
\end{table}

\clearpage
\begin{figure}
\centering
\psfig{figure=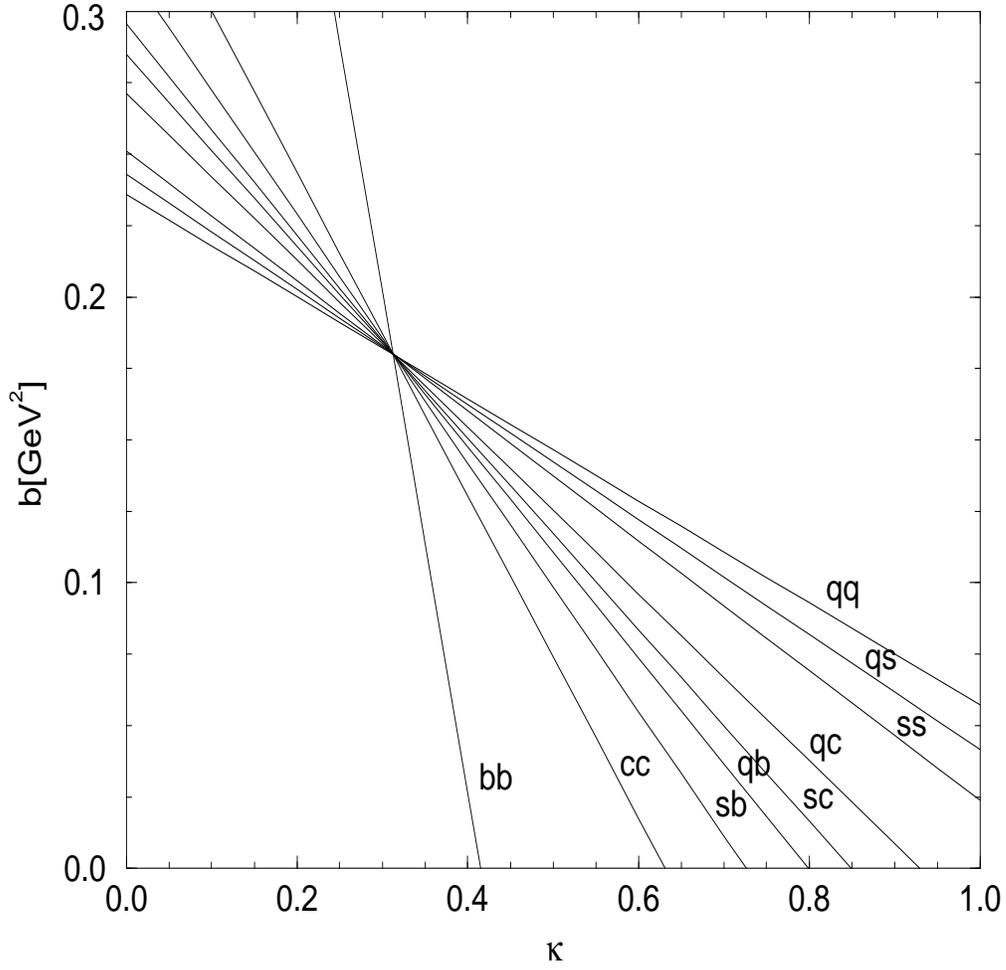,width=6in,height=6in}
\caption{The parameters $m_{s}$, $m_{c}$, $m_{b}$, $\beta_{qs}$, $\beta_{qc}$,
etc. satisfying variational principle.
The $qq$ and $qc$ etc. represents the sets of
$(m_{q},m_{q},\beta_{qq})$ and $(m_{q},m_{c},\beta_{qc})$ etc.,
respectively.}
\end{figure}
\setcounter{figure}{0}
\renewcommand{\thefigure}{\mbox{2.\alph{figure}}}
\begin{figure}
\centering
\psfig{figure=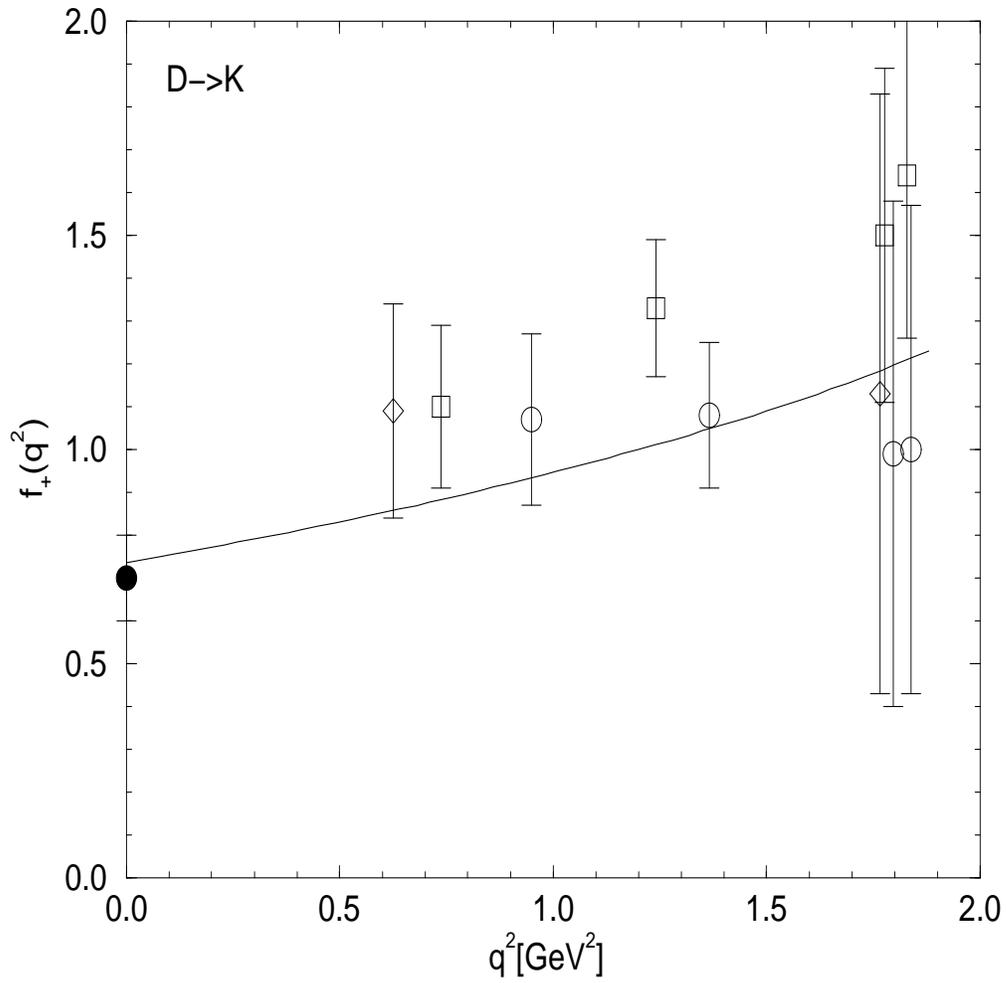,width=6in,height=6in}
\caption{The form factor $f_{+}(q^{2})$ for
$D$$\to$$K$ transition compared with the experimental data [1](full dot)
as well as the lattice QCD results [4].}
\end{figure}
\begin{figure}
\centering
\psfig{figure=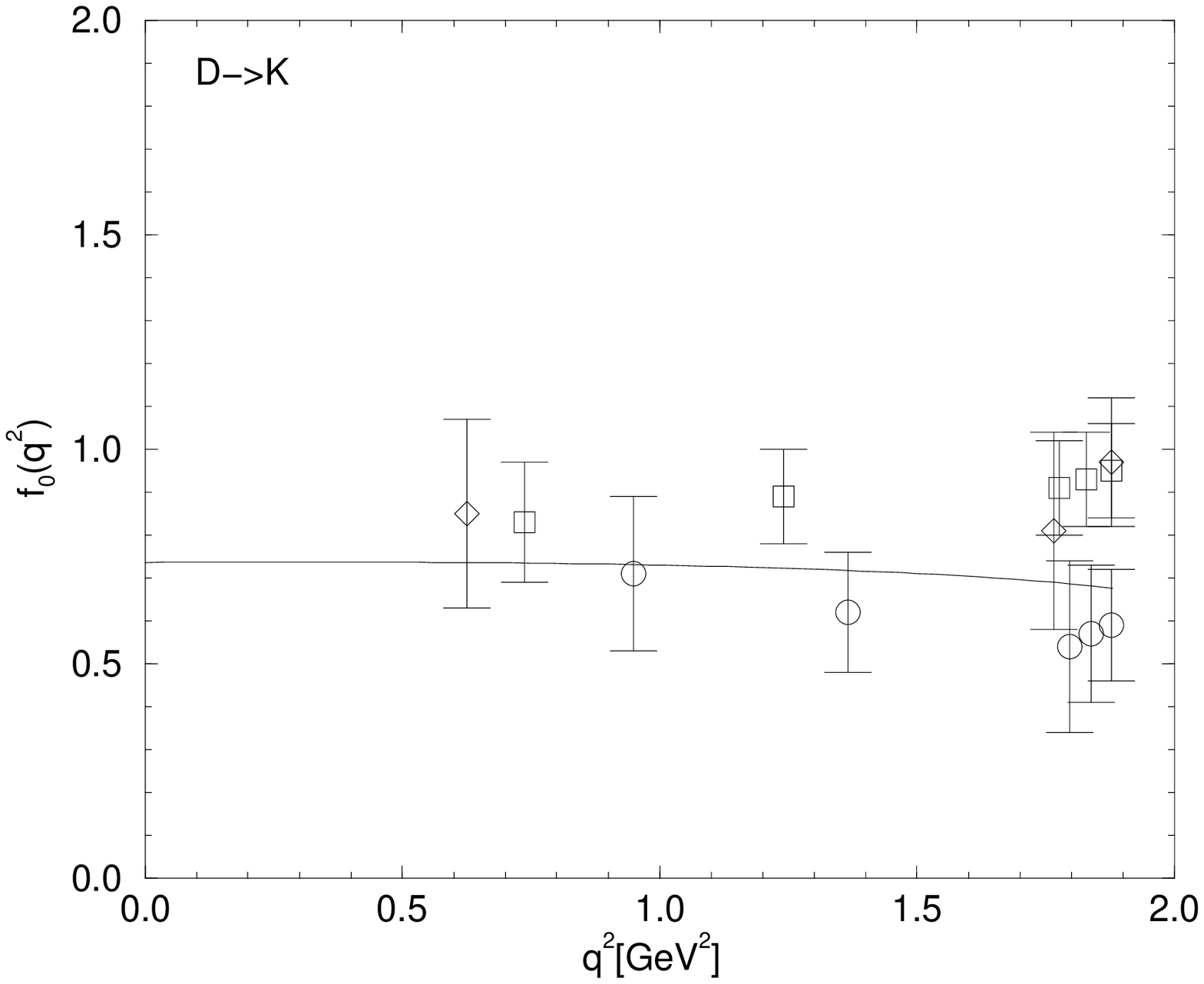,width=6in,height=6in}
\caption{The form factor $f_{0}(q^{2})$ for $D$$\to$$K$
transition compared with the lattice QCD results [4]. }
\end{figure}

\renewcommand{\thefigure}{\mbox{3}}
\begin{figure}
\psfig{figure=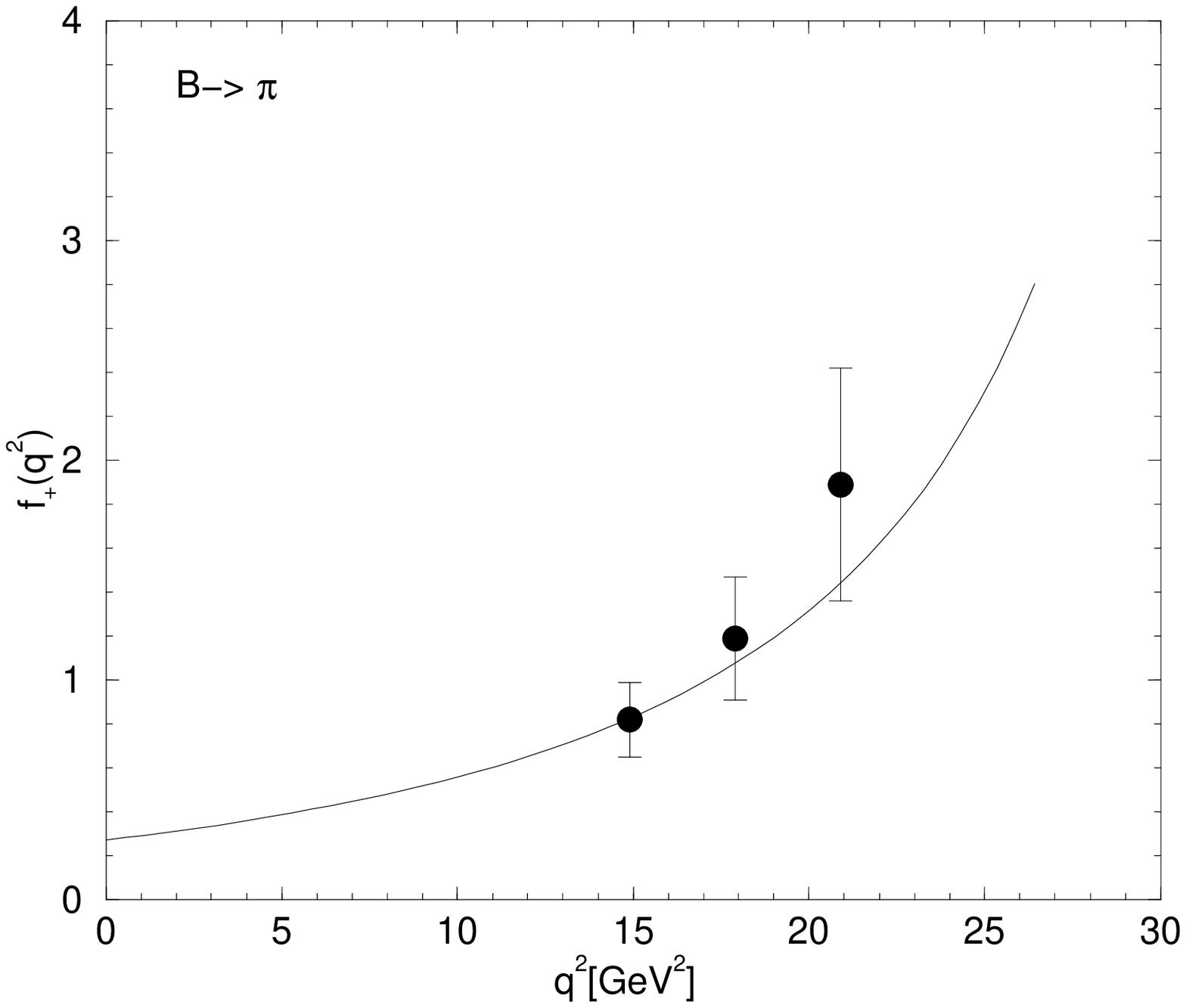,width=6in,height=6in}
\caption{ The form factor $f_{+}(q^{2})$ for
$B$$\to$$\pi$ transition compared with the
lattice QCD results [3].}
\end{figure}
\renewcommand{\thefigure}{\mbox{4}}
\begin{figure}
\psfig{figure=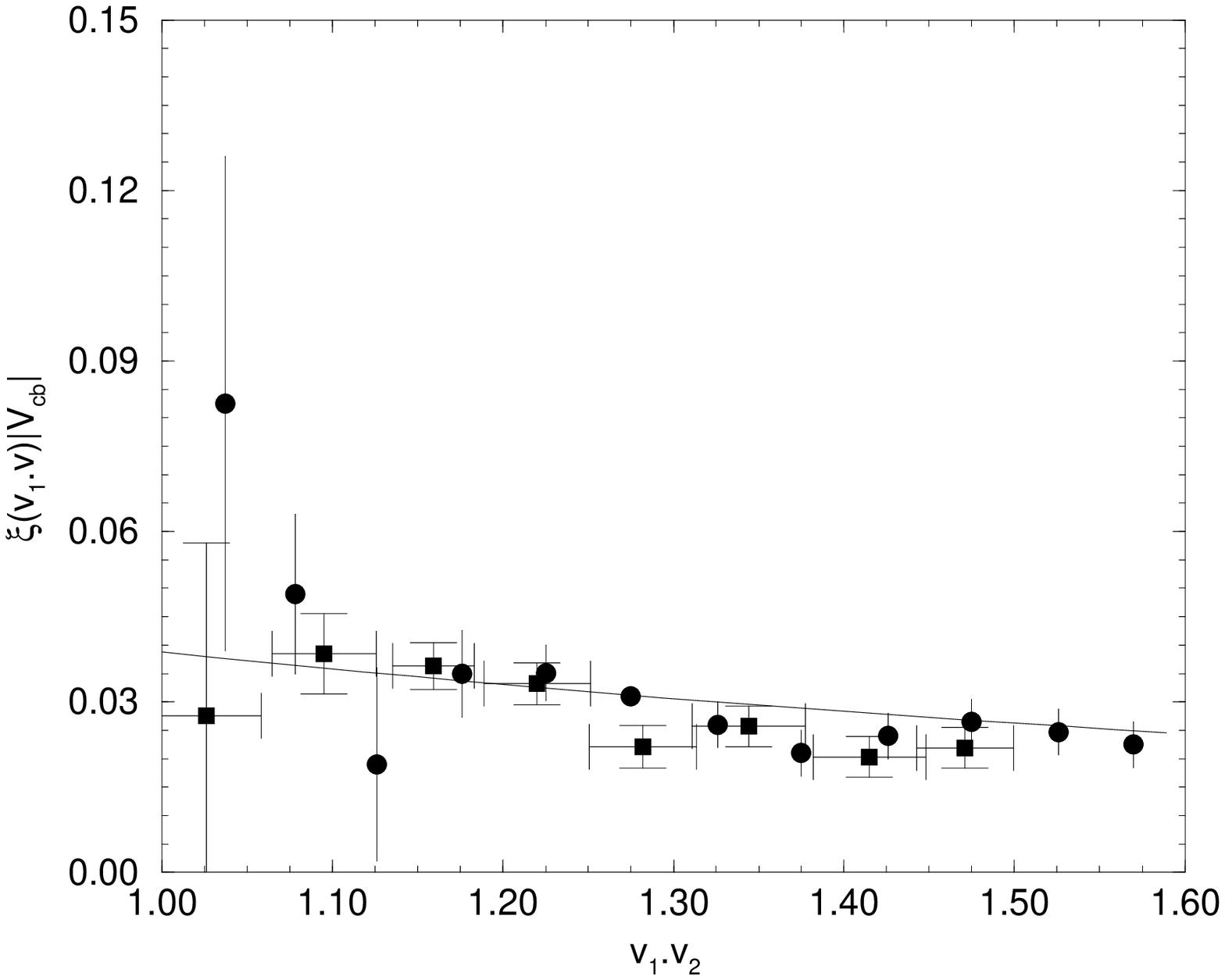,width=6in,height=6in}
\caption{ The IW function $\xi(v_{1}\cdot v_{2})$ for $B$$\to$$D$ transition
compared with the experimental data of ARGUS [21] (square) and
CLEO [22] (circle). }
\end{figure}

\end{document}